\documentclass[aps,pra,superscriptaddress,twocolumn,a4paper]{revtex4-1}
\usepackage[utf8]{inputenc}
\usepackage{amsmath}
\usepackage{graphicx}

\usepackage[left=2cm,right=2cm,
    top=2cm,bottom=2cm]{geometry}
\usepackage[english]{babel}

\usepackage{hyperref}
\hypersetup{
  pdfauthor = {Markus Sondermann},
  pdftitle = {Single photons from optically trapped nano-crystals},
}

\begin{document}

\title{Single photons emitted by nano-crystals optically
  trapped in a deep parabolic mirror}

\author{Vsevolod Salakhutdinov}
\affiliation{Friedrich-Alexander-Universit\"at Erlangen-N\"urnberg (FAU),
  Department of Physics, Staudtstr. 7/B2, 91058 Erlangen, Germany}
\affiliation{Max Planck Institute for the Science of Light,
Staudtstr. 2, 91058 Erlangen, Germany}

\author{Markus Sondermann}
\email{markus.sondermann@fau.de}
\affiliation{Friedrich-Alexander-Universit\"at Erlangen-N\"urnberg (FAU),
  Department of Physics, Staudtstr. 7/B2, 91058 Erlangen, Germany}
\affiliation{Max Planck Institute for the Science of Light,
Staudtstr. 2, 91058 Erlangen, Germany}

\author{Luigi Carbone}
\affiliation{CNR NANOTEC-Institute of Nanotechnology c/o Campus Ecotekne, University of Salento,
  via Monteroni, Lecce 73100, Italy}

\author{Elisabeth Giacobino}
\affiliation{Laboratoire Kastler Brossel, Sorbonne Université,
  CNRS, ENS-PSL, Research University, Collège de France, 4 place
  Jussieu,case 74 F-75005 Paris, France}
\affiliation{Max Planck Institute for the Science of Light,
Staudtstr. 2, 91058 Erlangen, Germany}

\author{Alberto Bramati}
\affiliation{Laboratoire Kastler Brossel, Sorbonne Université,
  CNRS, ENS-PSL, Research University, Collège de France, 4 place
  Jussieu,case 74 F-75005 Paris, France}

\author{Gerd Leuchs}
\affiliation{Max Planck Institute for the Science of Light,
Staudtstr. 2, 91058 Erlangen, Germany}
\affiliation{Friedrich-Alexander-Universit\"at Erlangen-N\"urnberg (FAU),
  Department of Physics, Staudtstr. 7/B2, 91058 Erlangen, Germany}
\affiliation{Department of Physics, University of Ottawa, Ottawa,
  Ont. K1N 6N5, Canada}

\begin{abstract}
We investigate the emission of single photons from
CdSe/CdS dot-in-rods which are optically trapped in
the focus of a deep parabolic mirror.
Thanks to this mirror, we are able to image almost the full 4$\pi$
emission pattern of nanometer-sized elementary dipoles and verify the
alignment of the rods within the optical trap.
From the motional dynamics of the emitters in the trap we
infer that the single-photon emission occurs from clusters comprising
several emitters.
We demonstrate the optical trapping of rod-shaped
quantum emitters in a configuration suitable for efficiently coupling
an ensemble of linear dipoles with the electromagnetic field in free space. 
\end{abstract}

\maketitle

\paragraph{Introduction}

Nano-scale light sources are used in many areas of research
and applications.
Examples range from the use of fluorescent nano-beads in microscopy to
the role of various nanometer-sized solid-state systems as sources of
single photons in future quantum technology applications.
One approach to efficiently collect the emitted
photons is to place the source at the focus of a parabolic mirror (PM) spanning
a huge fraction of the full solid angle.   
Such a mirror has been used to study a laser cooled atomic ion
held in a radio-frequency trap in vacuum~\cite{maiwald2012}.
More recently, the optical trapping of CdSe/CdS dot-in-rod
particles in a deep PM in air has been
demonstrated~\cite{Salakhutdinov2016}, and lately also the fabrication
of a microscopic diamond PM around a nitrogen vacancy
centre~\cite{wan2018}.

CdSe/CdS dot-in-rod (DR) particles have the remarkable properties of
being single-photon emitters at room temperature as well as emitting
light with a high degree of linear polarization~\cite{Pisanello2010}.
The latter property stems from the fact that their emission is dominated
by a linear-dipole transition~\cite{vezzoli2015}.
In order to maximize the collection efficiency when using a PM, the
axis of the linear dipole has to be aligned with the 
mirror's optical axis~\cite{Lindlein2007,sondermann2015}.
Once this is achieved, a PM with the geometry used in
Refs.~\cite{Lindlein2007,sondermann2015} as well as here collimates
94\% of the emission of a linear dipole, exceeding the capabilities
of a realistic lens-based set-up by a factor of two.
Moreover, the same requirement has to be fulfilled when reversing this scenario,
i.e. when focusing a linear-dipole mode with the purpose of exciting a
single quantum emitter with high efficiency~\cite{Lindlein2007}.
Any tilt of the quantization axis off the PM's axis will lead to a
reduced field amplitude along the transition dipole moment of the target.

For CdSe/CdS DR particles, the dipole-moment of the excitonic
transition is found to be aligned along the c-axis of the
rod~\cite{vezzoli2015,efros1996,mathieuThesis}.  Hence, the rod has to
be aligned along the optical axis of the PM for maximizing the
collection efficiency.  This should occur in a natural way in an
optical trap that is built by focusing a radially polarized beam as in
Ref.~\cite{Salakhutdinov2016}: The electric field in the focus of the
PM is purely longitudinal, i.e. parallel to the PM axis.  Since the
polarizability of a rod is largest along its symmetry
axis~\cite{Stickler2016}, the DR should thus be aligned along the
optical axis of the PM.  However, when trapping the particle under
ambient conditions, this alignment is counteracted by random
collisions with air molecules which exert a random torque on the DR
(see, e.g. Refs.~\cite{Hoang2016,Kuhn2017}). Thus, the torque induced
by the trapping laser onto the rod has to be larger than this random
torque in order to achieve a good alignment of the rods.

The purpose of this work is two-fold:
One aspect is checking the orientation of
the DRs with respect to the optical axis of the PM.
This is done by analyzing the spatially resolved pattern of the
photons emitted by CdSe/CdS DR-particles. 
Since the rods' alignment is governed by their geometry and
independent of the small CdSe core, our results should be applicable
for many types of rod-shaped quantum emitters.
In this sense, the CdSe/CdS DR-particles can be considered as a proof
model.
We again emphasize that a favorable alignment of a rod-shaped
quantum-system is a mandatory prerequisite for efficient photon
collection as well as for efficient excitation. 

The other aspect treated here is specific to the CdSe/CdS
DR-particles.
We will show that clusters of DRs can act
as a source of single photons.  
As a matter of fact, it will turn out that indeed all  
single-photon sources observed in our experiment are constituted by
clusters of emitters.
The emission of single photons from clusters of DRs demonstrated
here is in good agreement with recent observations of the same phenomenon
for CdSe/CdS core-shell dots~\cite{lv2018}.

\paragraph{Experimental procedure}
The CdS/CdSe DR-particles investigated here 
consist of a spherical CdSe core with $2.7\:\text{nm}$ diameter.
The core is embedded in a CdS rod with a nominal length of
$35\:\text{nm}$ and diameter of $7\:\text{nm}$.
Fluorescent photons are emitted at wavelengths about
$\lambda_\text{DR}=605\:\text{nm}$.
The DRs are trapped at ambient conditions in air at the focus
of a  deep parabolic mirror in an optical dipole-trap (see
Ref.~\cite{Salakhutdinov2016} and  Fig.~\ref{fig:setup} in appendix
for details). 
The electric field at wavelength $\lambda_\text{trap}=1064\:\text{nm}$
in the focal region is polarized along the optical axis of the
PM, which is here defined as the $z$-direction. 

At the beginning of each experiment the power of the trapping laser is
set to $360\:\text{mW}$ and the pulsed excitation beam is switched off. 
Successful trapping of DR particles is witnessed by the onset of a
fluorescence signal and from scattered trap light.
The fluorescence is detected with a pair of avalanche photo-diodes (APDs).
At this stage, fluorescence photons stem from two-photon
excitation at $\lambda_\text{trap}$~\cite{Salakhutdinov2016}.
To determine the intensity auto-correlation function $g^{(2)}(t=0)$, the power of the
trapping laser is reduced such that the count rate
of photons on the APDs is hidden in the background noise.
Then, more efficient pulsed excitation at $\lambda_\text{exc}=405\:\text{nm}$ is
switched on and the value of $g^{(2)}(t=0)$ is measured.

In case that $g^{(2)}(0)<0.5$ is observed, i.e. the signature of single-photon
emission, the experiment is continued as follows:
Images of the spatial intensity distributions of fluorescence photons
as found in the output aperture of the PM (see example in
Fig.~\ref{fig:fluorescence}) are acquired with an electron-multiplying
charge-coupled device (EMCCD) camera at several trapping beam powers.
For each power setting we have averaged over several images, with an
exposure time of $2\:\text{s}$ per image.
Vertically and horizontally polarized components are recorded
simultaneously on different parts of the camera chip. 

\begin{figure}[tb]
  \centering
   \includegraphics[width=8cm]{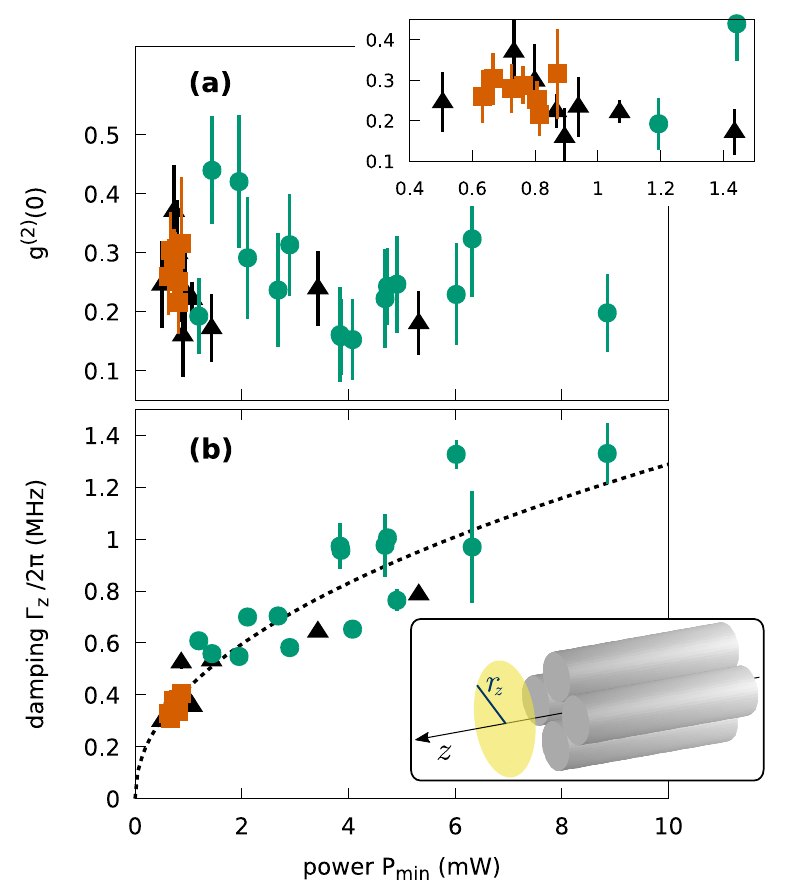}
   \caption{\label{fig:Pmin}
     (a) Normalized second-order intensity correlation $g^{(2)}(0)$ of
     fluorescence photons emitted by DRs against
     power $P_{min}$ at which the particles escaped the trap.
     Error bars show the Poissonian uncertainty.
     Green circles denote experiments in which the spatial intensity
     pattern revealed azimuthal symmetry and an obvious alignment of
     the rods to the axis of the PM. 
     A single such data point with $P_{min}=22\:\text{mW}$ and
     $g^{(2)}(0)=0.19$ is off scale.
     Orange squares correspond to cases where the emission pattern
     shows a clear azimuthal asymmetry.
     Black triangles mark experiments with an inconclusive outcome.
     The inset shows a magnified view for low power levels.
     (b) Damping rate of axial motion $\Gamma_z$ of trapped
     nano-particles against $P_\text{min}$. 
     Error bars denote the 95\% confidence interval of the fit from
     which $\Gamma_z$ was obtained.
     The dashed line denotes the function $\Gamma_z\propto
     P_\text{min}^{0.48\pm0.03}$, with the power scaling obtained from
     a fit.
     The inset is a pictorial view of a cluster of rods aligned along
     the optical axis of the PM ($z$-axis). The yellow disc with
     radius $r_z$ denotes the effective area of the cluster relevant
     for damping the motion along the $z$-axis.      
   }
\end{figure}

\begin{figure}
  \centering
  \includegraphics[width=8cm]{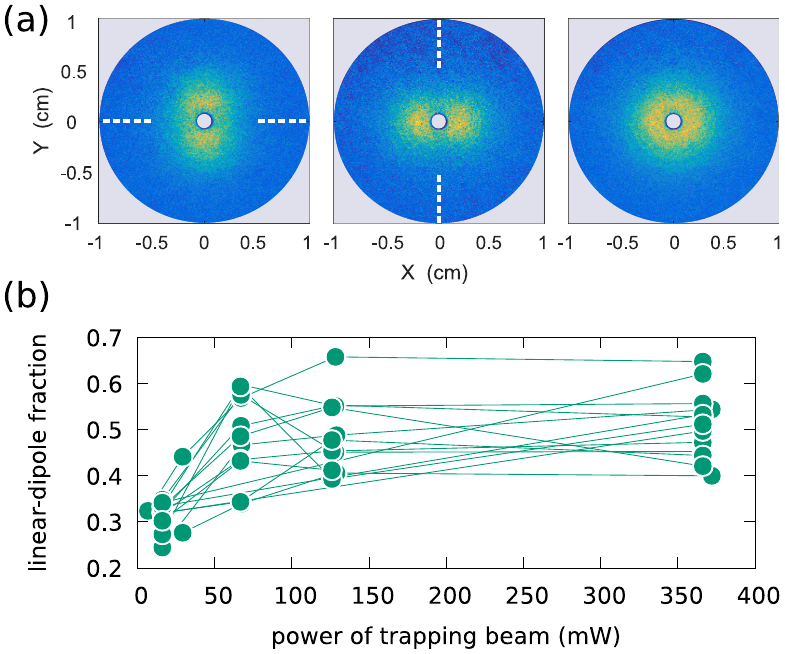}
  \caption{\label{fig:fluorescence}
    (a) Example of a spatial intensity pattern with rotational symmetry as
    observed in the aperture of the PM.  The corresponding sample is
    the one with $P_\text{min}=2.9\:\text{mW}$ and $g^{(2)}(0)=0.31$ in
    Fig.~\ref{fig:Pmin}.
    From left to right the panels show the intensity when projected
    onto a vertical state of polarization, the projection onto a
    horizontal state of polarization, and the sum of the two,
    respectively.
    Dashed lines indicate the orientation of extinction lines as
    a guide to the eye.
    (b) Fraction of linear-dipole radiation in the fluorescence of small
    clusters of DRs as a function of trapping beam power.
    Values obtained in the same experimental run are connected by
    lines.
    The data points correspond to the ones depicted by circles in
    Figs.~\ref{fig:Pmin}.
  }
\end{figure}

Next, the trapping beam power is set again to $360\:\text{mW}$
and the motional dynamics of the particles in the trap is analyzed.
The light scattered and Doppler-shifted by
the particle interferes with the trapping light~\cite{mestres2015}.
The corresponding interference signal is measured with a balanced
detector (cf. Ref.~\cite{Salakhutdinov2016}).
From the time series of this signal we compute the
power-spectral density which contains several Lorentzian-shaped
features, each of which corresponds to the motion of the particle in
the trap along a certain direction.
The width $\Gamma_i$ ($i=x,y,z$) of each of these peaks is proportional to the
ratio $r_i/m$, where $r_i$ is the radius of the trapped object and $m$ its
mass~\cite{neukirch2015r}.
Since the particles used here are non-spherical, we interpret $r_i$ as the
radius of a disc that has the same area as the particle's surface when
viewed from the direction of motion (see sketch in
Fig.~\ref{fig:Pmin}b).
With $m$ being proportional to the volume of the trapped object,
$\Gamma_i$ shrinks with increasing object size.
Fitting a Lorentzian function to the power spectral density yields the
axial damping rate $\Gamma_z$ of the particles'
motion~\cite{Salakhutdinov2016}. 
The motion in radial direction is damped such strong that a
corresponding spectral feature cannot be discerned. 

Finally, the trapping beam power is gradually decreased to determine
the power $P_\text{min}$ at which the particle is lost from the trap.
The loss of the trapped particles was monitored via the signal of the
balanced detector.

\paragraph{Single-photon emission from clusters}

In our experiments, we observed $g^{(2)}(0)<0.5$ for 33 different loads of
the trap, see Fig.~\ref{fig:Pmin}(a).
The probability to trap a sample with this characteristic is
approximately 10\%.
The observed values $0.15\le g^{(2)}(0)\le 0.44$ agree well with the ones
observed for single DRs with comparable geometry in 
literature~\cite{Vezzoli2013,manceau2014}.
Therefore, one could conclude that the trapped objects are indeed
single DRs.
But this conclusion is contradicted by the observation that all
trapped particles, which appear as single-photon emitters, are lost at
powers in the range $0.5\:\text{mW}<P_\text{min}<22\:\text{mW}$,
with an average value of $2.5\pm2.1\:\text{mW}$, see histogram in
Fig.~\ref{fig:sizeHist} in the appendix.
As derived there, the power at which the trapping
potential for a single DR equals $k_\text{B}T$ at ambient
temperature is $P_\text{min}\approx41\:\text{mW}$.
Therefore, a single DR that thermalizes to room temperature should be
lost at about this power value.
The volume and hence the polarizability of a trapped cluster
consisting of several rods scales with the number of rods.  
The magnitude of $P_\text{min}$ consequently reduces by the same factor.
Thus, one can infer that the smallest cluster displayed in
Fig.~\ref{fig:Pmin} must contain at least four DRs and that the
  average number of rods in a cluster is $16\pm14$.

It is also evident from the same figure that two branches 
in the scattered data points are apparent.
In one of these regions, which extends towards power values
$P_\text{min}\lesssim1.5\:\text{mW}$, the data points are densely
packed.
We will refer to these samples as \lq large clusters\rq.
At $P_\text{min}>1.5\:\text{mW}$ the data points are less dense.
These samples will be called \lq small clusters\rq.
The above threshold value of $P_\text{min}$ corresponds to approximately
27 rods.
Further below we show that these two branches are linked to
qualitative differences in the spatial intensity pattern of the
emitted photons, influenced by the geometry of the clusters and their
relative alignment to the PM axis. 
Independent of these differences, each branch has the property that
$g^{(2)}(0)$ tends to increase with decreasing $P_\text{min}$ or
increasing cluster size, respectively.  

The trapping of clusters is furthermore confirmed by analyzing the
damping rate of the particles' motion inside the trap.
Figure~\ref{fig:Pmin}(b) shows the axial damping rate $\Gamma_z$
as a function of $P_\text{min}$.
One observes a clear correlation of the two quantities with
$\Gamma_z\propto \sqrt{P_\text{min}}$.
Since $\Gamma_z$ increases with a decreasing size of the trapped
object, we conclude that the variation in $P_\text{min}$ is due to the
variation of the size and, therefore, to the polarizability of the
trapped object. 
Furthermore, all observed damping rates are smaller than the one
estimated for a single DR ($\Gamma_{1\text{DR}}/2\pi\ge
2\:\text{MHz}$, see appendix).
Hence, the trapped emitters of different size must be clusters
consisting of differing numbers of DRs. 

Surprisingly, the trapping of DRs in clusters occurs despite the use of
toluene as working solvent that should suppress the formation of
clusters in the aerosol sprayed in the trapping region. Possibly,
clusters might form inside the optical trap as soon as several
single DRs diffuse in the focal region of the PM.

The observation of pronounced antibunching in the emission from a
cluster of sources proves the emission of single photons.
This is not to be expected at first sight, however, but in good
agreement with findings in other experiments.
In Ref.~\cite{goodfellow2016} the quenching of the emission of
CdSe/CdS core-shell quantum-dots by Förster energy-transfer to a
$\text{MoSe}_2$ mono layer was demonstrated, which has earlier
been also observed within assemblies of CdSe
dots~\cite{kagan1996,crooker2002}.
In principle, such an exchange could also occur between several DRs in
a cluster provided that the spectra of excitonic transitions of the
DRs constituting the cluster overlap.
Recently, single-photon emission from clusters of CdSe/CdS core-shell
dots has been attributed to Auger annihilation of excitons in
neighboring nano-crystals~\cite{lv2018}.  
For DRs trapped sufficiently close to each other the same mechanism
might be responsible for single-photon emission.

\paragraph{Analysis of the fluorescence pattern}

We now investigate the spatial intensity pattern of the photons
emitted by the DRs.
Since the polarizability of a rod is larger along its symmetry axis
than perpendicular to it, one expects that a single rod or a small
enough cluster of parallel rods is oriented along the longitudinal
electric field in the focus of the PM.
Under this assumption and with the dipole moment of the exciton
aligned along the rod, the intensity distribution of the photons
collimated by the PM has to be symmetric under rotation around the
optical axis of the PM. 
In the majority of loads (26 out of 33) the measured
distribution of the total intensity exhibits this symmetry. 

While a rotationally symmetric intensity pattern could in principle also be
observed for an isotropic emitter, more insight can be gained from
the polarization resolved emission pattern.
For a linear dipole aligned parallel to the axis of the PM, the
collimated emission is radially polarized~\cite{Lindlein2007}.
After projection onto a linear state of polarization, the intensity
distribution of a radially polarized state must exhibit extinction
lines perpendicular to the projected polarization.
This property is illustrated in the example displayed in
Fig.~\ref{fig:fluorescence}(a).
The observed extinction is not perfect, which is compatible with the
existence of circular-dipole components in the DR emission. 

The clusters which emit photons with a rotationally symmetric
spatial distribution and which exhibit polarization patterns as
described above are analyzed in further detail:
We compute the azimuthal average of the total intensity
distribution, resulting in a radial intensity profile~\cite{maiwald2012}.
Fitting a sum of the intensity distributions of linear and circular
dipoles~\cite{Sondermann2008} to these profiles reveals the
\emph{measured} fraction of linear-dipole radiation (see App.~\ref{sec:dipole}).
This measured fraction will be lower than the one predicted by the
properties of the excitonic transitions of the emitters due to the
residual thermal vibration of the rods.
The latter results in randomly fluctuating tilts of the rod
off the PM's axis.
With tilts into all directions being equally likely,
a linear dipole appears not as a pure linear dipole but is
masked by a circular-dipole component in the long-time
average.
However, the thermal vibrations are counteracted by the longitudinal
electric field of the trap.
This field exerts a torque onto non-aligned rods and
re-orients them along the PM axis. 
Therefore, a stronger suppression of
random tilts should occur at larger field amplitudes, i.e. at larger
trapping beam powers. 
One thus expects to observe a larger linear-dipole fraction at larger
trapping beam powers. 

Indeed, 16 samples exhibit such a behavior, as is evident from
Fig.~\ref{fig:fluorescence}(b).
This confirms the alignment of the rods along the PM axis on
average. 
Besides one exception, all samples (marked with circles in
Figs.~\ref{fig:Pmin}) showing axial alignment were
lost from the trap at  $P_\text{min}>1.5\:\text{mW}$. 
These samples also constitute the group with the largest damping rates.
Therefore, we conclude that the samples lost at
$P_\text{min}>1.5\:\text{mW}$ are small clusters of DRs
that are aligned along the axial trapping field.
It should be noted that there is no correlation between the
linear-dipole fraction measured at large trapping beam power and the
power $P_\text{min}$ at which samples are lost from the trap.
Alignment along the field of an optical trapping beam was previously
observed also for silicon nanorods~\cite{Kuhn2017} and
nano-diamonds~\cite{Hoang2016,Geiselmann2013} as well as for nanorods
trapped in a liquid~\cite{ruijgrok2011,head2012}.

Most of the samples lost at low powers 
show no uniquely classifiable behavior in terms of emission pattern.
This hints at emission from a large cluster without a
pronounced structural order.
Some of these large clusters show an emission pattern without
rotational symmetry, i.e. with the dipole axis tilted off the PM's
optical axis.
We conjecture that these cases correspond to emission from
DRs that are attached to a large main cluster under a non-zero angle.
However, once released from the trap the clusters are
inevitably lost, hindering a posteriori investigations of the
clusters' structure.
 
A tempting explanation why the majority of aligned samples is found
among the small clusters can be constructed from a geometric argument as
follows:
If the rods are aligned parallel to each other in closest packing,
the width of the cluster equals the length of a single rod for a
number of about 20 rods, resulting in an object of approximately equal
size along all spatial dimensions and leading to the loss of
pronounced directionality.   
In the experiments the apparent change from the alignment of clusters
along the PM axis to no preferential alignment occurs at a number of
about 27 rods.
This is in reasonable agreement with the number inferred just above.
But the fact that some of the large clusters exhibit radiation
patterns lacking rotational symmetry hints at a more complicated scenario. 

Finally, it would be interesting to compare the measured count rates
of photons emitted by the DRs to the ones expected for our set-up,
in particular for the samples exhibiting alignment.
However, a reliable quantitative comparison is hindered by the fact that the
investigated samples exhibit blinking,  as detailed in App.~\ref{sec:countRates}.

\paragraph{Concluding discussion}

We have analyzed the fluorescence photons emitted by CdSe/CdS DRs which are
optically trapped at the focus of a deep PM.
The observation of single-photon emission from clusters of sources
will be advantageous for applications, since clusters can be kept in
the trap at trapping beam powers that are lower than necessary for
single emitters.
Moreover, for low trapping beam powers the excitation of the DRs via
two-photon absorption~\cite{Salakhutdinov2016} can be strongly
suppressed, which promises a better timing accuracy for the
emission of photons in a pulsed excitation scheme.

On the one hand, it could be possibly argued that the trapping of
clusters of DRs, i.e. the fact that so far not a single isolated DR
was trapped, might be detrimental to the efficient coupling of light
to these emitters.  
On the other hand, this appears as a favorable condition to possibly
promote the enhancement of the coupling due to collective effects when
trapping a cluster of quantum targets.  
Also in cavity quantum-electrodynamics collective effects have been
used
advantageously~\cite{thompson1992,lambrecht1996,brennecke2007,colombe2007,herskind2009}.
There, large fractions of the ensemble can be hosted in the cavity
mode~\cite{brennecke2007,colombe2007}. 
When coupling light and ensembles in free-space by tight focusing, the
ensemble has to be located within the focal spot, which is of
dimensions of the order of half a wavelength under optimum
conditions~\cite{gonoskov2012,alber2017}.
This is very difficult to achieve with ions due to Coulomb repulsion.
Therefore, locating small clusters of optically trapped
solid-state quantum targets in the focus of a linear-dipole wave is a
feasible path towards combining tight focusing and ensemble physics.

\begin{acknowledgments}
The authors thank M.\,Manceau for fruitful discussions.
G.L. acknowledges financial support from the European Research Council
(ERC) via the Advanced Grant 'PACART'. 
\end{acknowledgments}

%

\onecolumngrid
\clearpage

\appendix

\section{Experimental set-up}

\label{app:setup}

\begin{figure}[b]
  \centering
   \includegraphics[width=8cm]{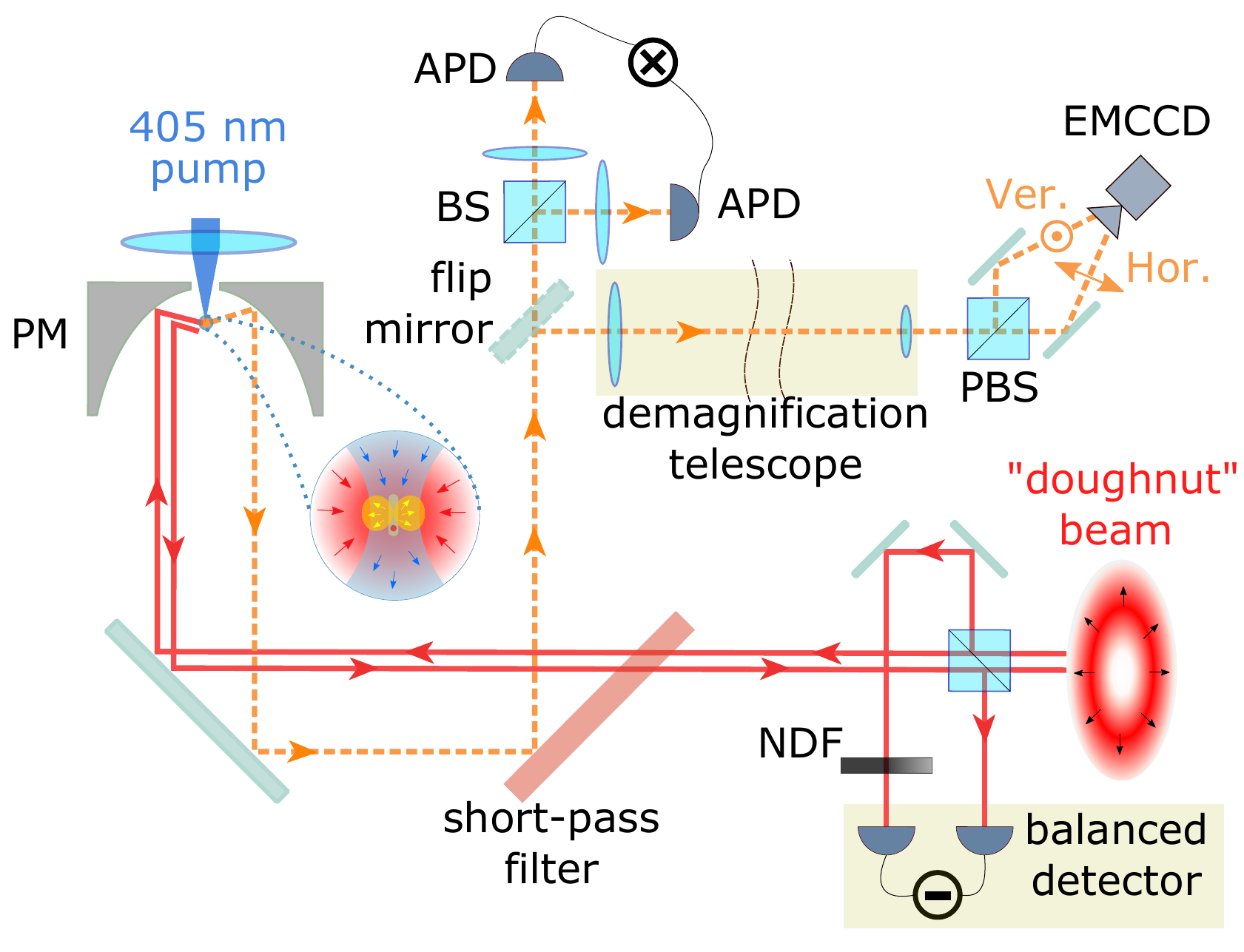}
   \caption{\label{fig:setup}
     Scheme of the experimental set-up.
     CdSe/CdS dot-in-rod (DR) particles are trapped in an optical tweezers by
     focusing a radially polarized infrared laser beam (\lq doughnut beam\rq)
     with a parabolic mirror (PM).
     The DRs are optically excited by a blue pulsed laser (\lq pump\rq).
     Fluorescent photons are collimated by the PM
     and either detected by two avalanche photo-diodes (APDs), which
     are separated by a beam splitter (BS), or an
     electron-multiplying charge-coupled device (EMCCD) camera.     
     A polarizing beam splitter (PBS) in front of the EMCCD camera
     projects the photons onto horizontally and vertically polarized
     states measured by different sections of the camera chip.
     The motion of the DRs is characterized by 
     detection of the light of the trap laser that is scattered by the 
     DRs and interferes with trap light back-reflected by the
     PM. This signal is balanced with light that is tapped off the trap laser beam
     and adjusted in amplitude by a neutral-density filter (NDF).
     } 
\end{figure}

Figure~\ref{fig:setup} shows a schematic of the experimental set-up.
Nano-particles are trapped at ambient conditions in air at the focus
of a  deep parabolic mirror (PM).
The PM has a focal length of $2.1\:\text{mm}$ and an aperture radius
of $10\:\text{mm}$, which corresponds to a half-opening angle of
$134^\circ$.
A bore hole with $1.5\:\text{mm}$ diameter at the vertex of the
PM allows for the delivery and the optical excitation of the
nanoparticles.
Based on these geometrical boundary conditions the collection
efficiency for photons in a linear-dipole mode (circular-dipole
mode) cannot exceed $94\%$ ($76\%$), where the quantization axis is
defined to coincide with the optical axis of the PM.
The reflectivity of the PM was determined to be $72\%$.

The trapping beam is radially polarized and has the intensity pattern
of a Laguerre-Gaussian mode of first radial order (\lq doughnut mode\rq).
The electric field at wavelength $\lambda_\text{trap}=1064\:\text{nm}$
in the focal region is polarized along the optical axis of the
PM, which is here defined as the $z$-direction. 

The CdS/CdSe dot-in-rod (DR) particles are surrounded by alkyl chains stored in toluene,
hindering the clustering of particles.  
We use a home-built, toluene-resistant nebulizer for
delivering DR particles to the trapping region through an opening at the
vertex of the PM.
The nebulizer comprises a piezo-electric transducer that is driven
with an alternating voltage oscillating at about $360 - 390\:\text{kHz}$.
For these frequencies one expects an aerosol of toluene droplets with a mean
diameter of $6\:\mu\text{m}$~\cite{Lang1962, Barreras2002}.
The nebulizer is filled with toluene containing DR particles in
concentrations ranging from $6\times10^{4}$ to
$6\times10^{8}\,\text{DRs}/\text{l}$.
We thus expect less than $10^{-4}$ DRs in a single droplet of about
100\,fl volume on average.

Light scattered and Doppler-shifted by
the particle interferes with the trapping light~\cite{mestres2015}.
The corresponding interference signal is measured with a balanced
detector (cf. Fig.~\ref{fig:setup} and Ref.~\cite{Salakhutdinov2016}).
The balancing is achieved with light that is tapped off the trap laser beam
and adjusted in amplitude by a neutral-density filter (NDF).

Trapped DR particles are optically excited at a wavelength
$\lambda_\text{exc}=405\:\text{nm}$ with pulses of $82\:\text{ns}$
duration and $2\:\mu\text{W}$ average power. The pulse repetition rate
is $10^6\:\text{s}^{-1}$. 
The excitation pulse power is chosen such that the DRs are
approximately saturated, i.e. on average one exciton is created per
excitation pulse~\cite{Vezzoli2013}.
The DRs emit photons at wavelengths about $\lambda_\text{DR}=605\:\text{nm}$,
which are picked off the excitation and trapping beam path by
a reflective short-pass filter that attenuates light at
$\lambda_\text{trap}$ in reflection.
The fluorescence photons are analyzed by either a pair of
APDs with $69\%$ quantum efficiency for determining the
second-order intensity correlation function 
$g^{(2)}(t)$ or by an EMCCD camera for measuring the spatial intensity
distribution as found in the output aperture of the PM.
This aperture is imaged onto the camera by a demagnifying 
telescope.
Vertically and horizontally polarized components are split by a
polarizing beam splitter (PBS) and recorded
simultaneously on different parts of the camera chip. 
The APDs as well as the camera are equipped with band-pass filters
which reject light at $\lambda_\text{exc}$ and at
$\lambda_\text{trap}$.
Not accounting for the reflectivity of the PM as well as the
APDs' quantum efficiency, the transmission of the set-up was
measured to be $83\%$.

\section{Minimum trapping beam power}
\label{app:trap}

The depth of the trapping potential of a dielectric particle in the
Rayleigh approximation is given by $U_0=\alpha/2\cdot E_\text{max}^2$,
where $\alpha$ is the polarizability of the particle and
$E_\text{max}$ is the maximum amplitude of the electric field in the
focal region.
$E_\text{max}^2$ is proportional to the power of the focused trapping
beam.
The derivation of $E_\text{max}$ for our particular set-up can be found
in Ref.~\cite{Salakhutdinov2016}.
Taking the polarizability $\alpha$ to be the one of a CdS rod
aligned to the electric field, one has~\cite{Kuhn2017}
\begin{equation}
  \label{eq:alpha}
  \alpha=\epsilon_0 V_\text{rod}(n_\text{CdS}^2-1)
\end{equation}
with $n_\text{CdS}=2.34$ the
refractive index of CdS at $\lambda_\text{trap}$ and
$V_\text{rod}$ the volume of the rod.
Setting $U_0=k_\text{B}T$ with $T=296\:\text{K}$ yields
$P_\text{min}=41\:\text{mW}$.

\begin{figure}
  \includegraphics{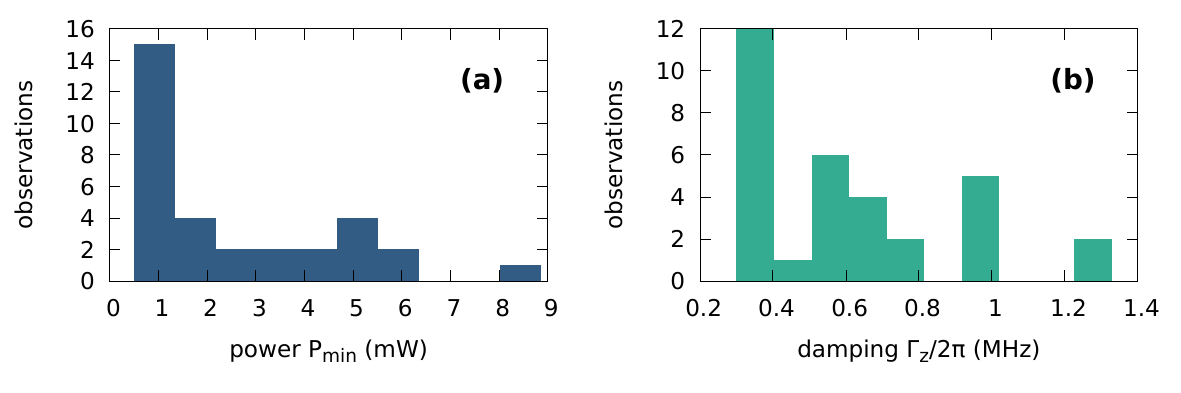}
  \caption{\label{fig:sizeHist}
    (a) Distribution of minimum trapping beam powers $P_\text{min}$
    at which samples escaped the trap.
    (b) Distribution of damping rates $\Gamma_z$ of the axial motion
    of the samples trapped in the experiment.
    The histograms have been computed for the data shown in Fig.~\ref{fig:Pmin} in
    the main text. 
  }
\end{figure}

The distribution of powers $P_\text{min}$ at which samples with
$g^{(2)}(0)<0.5$ escaped the trap is shown in Fig.~\ref{fig:sizeHist}.
The average value (standard deviation) is $2.5(2.1)\:\text{mW}$.
This corresponds to an average number of $16(14)$ rods in a trapped
cluster.

\section{Damping of axial motion}

For a spherical particle in air, the damping of its oscillatory motion
is given by~\cite{neukirch2015r}
\begin{equation}
  \label{eq:damping}
  \Gamma=\frac{6\pi\eta r}{m}
  \cdot\frac{0.619}{0.619+\text{Kn}}(1+c_\text{K})\quad ,
\end{equation}
with $\eta$ the viscosity of air, $r$ the sphere's radius and $m$
its mass.
$\text{Kn}=\Lambda/r$ is the Knudsen number with $\Lambda$ the mean
free-path length of an air molecule and
$c_\text{K}=0.31\text{Kn}/(0.785 + 1.152\text{Kn} + \text{Kn}^2)$.

Here, we interpret $r$ as the radius of the area of the particle's
surface as seen along the direction of motion. 
Hence, for a rod aligned parallel to the optical axis of the PM $r$ is
identified with the radius of the rod.
Due to the attached alkyl chains, which have a length of
about $1.6\:\text{nm}$, the \textit{effective} radius of the rod is
increased, resulting in $\Gamma_\text{1DR}/2\pi\approx
2\:\text{MHz}$.
Furthermore, due to residual vibration the
trapped particles' effective radius is enlarged when
averaging over the time constants of translational motion.   
The distribution of damping rates observed in the experiments for
samples with $g^{(2)}(0)<0.5$ is shown in Fig.~\ref{fig:sizeHist}. 
The average value (standard deviation) of $\Gamma_z/2\pi$ is
$0.62(0.29)\:\text{MHz}$.

\section{Dipole radiation patterns}
\label{sec:dipole}

The analysis of the spatial intensity pattern of the photons
emitted by the DRs is based on determining the relative fraction of
linear- and circular-dipole radiation.
We choose the quantization axis to coincide with the optical axis of
the PM.
Then, upon collimation by the PM the intensity distribution of a
linear dipole reads~\cite{Lindlein2007,Sondermann2008}
\begin{equation}
  I_\pi(R)=I_{0,\pi}\cdot\frac{R^2}{(R^2/4+1)^4}
\end{equation}
with $I_{0,\pi}$ a proportionality constant and $R$ the radial distance to
the optical axis of the PM in units of the mirror's focal length.
The corresponding expression for a circular dipole reads
\begin{equation}
  I_\sigma(R)=I_{0,\sigma}\cdot\frac{R^4/16 + 1}{(R^2/4+1)^4}
  \quad .
\end{equation}
The sum of these two distributions is fit to radial intensity
profiles which are obtained by computing the azimuthal average of the
images acquired with the EMCCD camera. 
The resulting fraction of linear-dipole radiation is then given by
\begin{equation}
  A_\pi=\frac{I_{0,\pi}}{I_{0,\pi}+I_{0,\sigma}} \quad .
\end{equation}

\section{Photon count rates}
\label{sec:countRates}

In what follows, we first derive the photon count rate expected for
the CdSe/CdS DRs and then compare to the ones observed in the experiment.
One has to account for several characteristics of the investigated emitters.
The number of electron-hole pairs created upon optical excitation with
power $P$ is given by $P/P_\text{sat}$, where $P_\text{sat}$ is the
saturation power defined as the power at which one exciton-hole pair
is created on average~\cite{Vezzoli2013}.
The number of emitted photons is proportional to
$(1-\exp(-P/P_\text{sat}))$. 
We have determined $P_\text{sat}$ from saturation
measurements~\cite{Vezzoli2013} using large ensembles of rods,
i.e. clusters with intensity correlation $g^{(2)}(0)\approx 1$.
The resulting value is $P_\text{sat}=2.63\pm0.43\:\mu\text{W}$.
The used excitation power in all our experiments was
$P=2\:\mu\text{W}$.
Furthermore, the DRs have a non-unit quantum yield. The latter was
previously determined to be about $Q_\text{DR}=0.7$~\cite{carbone2007}.

In the experiment the photons are collected upon pulsed excitation
from the auxiliary laser at a rate $\gamma_\text{exc}=10^6/\text{s}$
and at strongly reduced trapping beam power.
The latter condition results in an ineffective suppression of rod
vibrations, such that the \emph{measured} radiation pattern contains a
fraction of linear-dipole radiation of $A_\pi=0.31\pm0.03$ (cf. Fig.~\ref{fig:fluorescence}
in main text).
The collection efficiency of the used PM itself is $94\%$ for a linear
dipole and $76\%$ for circular dipole radiation.

Taking into account the APD's quantum efficiency $Q_\text{APD}$, the
reflectivity $R_\text{PM}$ of the PM, and the transmission $T$ through the setup
(all specified in Sec.~\ref{app:setup}), the expected photon count rate is
\begin{equation}
  \gamma_\text{exc}\cdot Q_\text{APD} \cdot T \cdot R_\text{PM}\cdot [1-\exp(-P/P_\text{sat})]
  \cdot Q_\text{DR} \cdot [0.94\times A_\pi +0.76 \times (1-A_\pi)] =
  (125\pm 14)\times 10^3\,\text{s}^{-1}\quad .
\end{equation}

\begin{figure}
  \centering
  \includegraphics{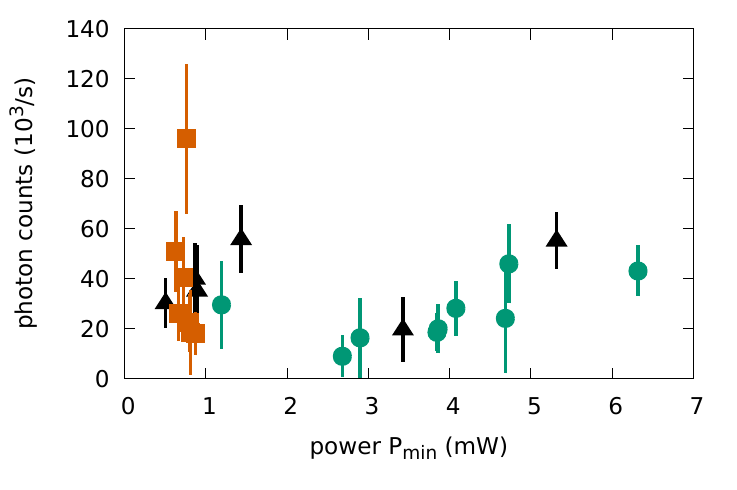}
  \caption{\label{fig:counts}
    Photon count rates of trapped clusters as a function of
    power $P_\text{min}$ at which samples escaped the trap.
    The count rates correspond to the average rate of a grey state
    obtained by fitting a gaussian distribution to count rate
    histograms.
    The errorbars represent the rms width of the gaussian distribution.
    The meaning of different symbols is the same as in
    Fig.~\ref{fig:Pmin} in the main text.
    }
\end{figure}

All samples investigated in the experiment show a blinking behavior
which is well-known in literature.
Therefore, for the analysis of the photon detection events we apply
established methods accounting for this
effect~\cite{spinicelli2009,pisanello2013,manceau2014}.
The time-resolved count rate of detected photons is
obtained by binning the time-tagged detection events into a continuous
series of intervals with a constant width of $500\,\mu\text{s}$.
From the resulting time series the distribution of count rates is
computed.
For some samples, the time series is dominated by intermittent bursts from a low
light level.
The corresponding histograms show an exponentially decaying distribution of count
rates.
For 22 samples the distribution of count rates exhibits a pronounced local
maximum at nonzero detection rates.
The average count rates corresponding to this local maximum are
reported in Fig.~\ref{fig:counts}.
The observed rates are considerably below the expected rate.

However, the observed count rates are strongly fluctuating from
sample to sample.
The size of the cluster, i.e. the value of $P_\text{min}$  has no
obvious influence onto the rate of detected photons. 
One might argue that there is a slight tendency of increasing count
rate for increasing $P_\text{min}$ evident from Fig.~\ref{fig:counts}
for small clusters with $P_\text{min}>1.5\,\text{mW}$, i.e. the
aligned samples.
But the corresponding count rates are of comparable magnitude to the
rates observed for the large samples.
Furthermore, the photon emission properties of all samples are
investigated at low trapping beam powers, at which the alignment of
the small clusters is weakest anyhow. 
There is also no evident correlation of count rate and the value of
$g^{(2)}(0)$ (data not shown).

These observations indicate that the photon count rates are related to the
photo-physics of the DRs and not the properties of the
setup or the geomnetry of the trapped clusters.
Furthermore, the blinking behaviour is typically characterized by
distributions exhibiting two pronounced maxima at nonzero count
rates (cf. e.g. Refs.~\cite{spinicelli2009,manceau2014}), where the
larger rate corresponds to a so-called \lq bright-state\rq\ and the lower one
to a so-called \lq grey state\rq\ which is related to the presence of
excess charges~\cite{galland2011}.
None of the samples investigated here exhibits two pronounced states.
Due to the low photon count rates we associate the rates given in
Fig.~\ref{fig:counts} with a grey state.
In addition, the emission from a grey state is typically a factor 2 to
6 weaker than from a bright state~\cite{spinicelli2009,manceau2014}.  
When applying this correction factor, many samples would exhibit an
extrapolated bright state with count rates in reasonable agreement
with the expected one. 
We attribute the dominance of emission from a grey state to
particulary strong charging of the optically trapped samples.
A potential origin of the charging is the process of spraying the
aerosol containing the nano-particles into the trapping region.
For example, also silica nano-spheres are typically charged after
spraying~\cite{moore2014,frimmer2017}.

\end{document}